%% file: dsqnm.tex
\documentclass[aps,twocolumn,showpacs,preprintnumbers,nofootinbib,prd,superscriptaddress,groupedaddress,10pt]{revtex4-1}

\makeatletter
\def\l@subsubsection#1#2{}
\def\l@subsubsubsection#1#2{}
\makeatother

\usepackage{graphicx,amssymb,amsmath,amsthm,amsfonts,epsfig,epsf}
\usepackage[usenames]{color}
\usepackage{epstopdf}
\definecolor{darkred}{rgb}{0.5,0,0}

\usepackage{array}
\usepackage{afterpage}
\usepackage{bm}
\usepackage{dcolumn}
\usepackage[latin1]{inputenc}
\usepackage{latexsym}
\usepackage{rotating}
\usepackage{longtable}

\usepackage{enumerate}
\usepackage{tensor,multirow}
\usepackage{url}
\usepackage{xcolor}
\definecolor{winered}{rgb}{0.6,0,0}
\definecolor{lessblue}{rgb}{0,0,0.7}
\usepackage[linktocpage,pdftex,colorlinks=true,linkcolor=winered,citecolor=lessblue,urlcolor=lessblue,breaklinks=true,bookmarksopen=true]{hyperref}

\usepackage{float}
\usepackage{graphicx}
\usepackage{colortbl}

\newcommand{\C}{\mathbb{C}}
\newcommand{\R}{\mathbb{R}}
\newcommand{\N}{\mathbb{N}}

\newcommand{\dd}{{\mathrm{d}}}
\newcommand{\pa}{\partial}

\newcounter{mnotecount}[section]

\renewcommand{\themnotecount}{\thesection.\arabic{mnotecount}}

\newcommand{\mnote}[1]%{}%
{\protect{\stepcounter{mnotecount}}$^{\mbox{\footnotesize
$%\!\!\!\!\!\!\,
\bullet$\themnotecount}}$ \marginpar{%\color{red}%
\raggedright\tiny\em
$\!\!\!\!\!\!\,\bullet$\themnotecount: #1} }

\begin{document}

\title{Quasinormal modes and dual resonant states on de~Sitter space}

\author{
Peter Hintz$^{1}$,
YuQing Xie$^{2}$
}
\affiliation{$^1$ Department of Mathematics, ETH Z\"urich, R\"amistrasse 101, 8092 Z\"urich, Switzerland}
\affiliation{$^1$ Department of Mathematics, Massachusetts Institute of Technology, Cambridge, MA 02139-4307, USA}
\affiliation{$^2$ Department of Physics, Massachusetts Institute of Technology, Cambridge, MA 02139-4307, USA}
\begin{abstract}
  The existence of quasinormal modes (QNMs) for waves propagating on pure de~Sitter space has been called into question in several works. We definitively prove the existence of quasinormal modes for massless and massive scalar fields in all dimensions and for all scalar field masses, and present a simple method for the explicit calculation of QNMs and the corresponding mode solutions. By passing to coordinates which are regular at the cosmological horizon, we demonstrate that certain QNMs only appear in the QNM expansion of the field when the initial data do not vanish near the cosmological horizon. The key objects in the argument are dual resonant states. These are distributional mode solutions of the adjoint field equation satisfying a generalized incoming condition at the horizon, and they characterize the amplitudes with which QNMs contribute to the QNM expansion of the field.
\end{abstract}

\maketitle
%

%%%%%%%%%%%%%%%%%%%%%%%%%%%%%%%%%%%%%%%%%%%%%%%%%%%%%%%%%%%%%%%%%%%%%%%%%%%%%%
\section{INTRODUCTION AND SUMMARY}
%%%%%%%%%%%%%%%%%%%%%%%%%%%%%%%%%%%%%%%%%%%%%%%%%%%%%%%%%%%%%%%%%%%%%%%%%%%%%%

The de~Sitter (dS) universe is the simplest solution of the Einstein vacuum equations with cosmological constant $\Lambda>0$. This makes it an ideal starting point for the investigation of a large variety of physical phenomena in universes, such as our own \cite{RiessEtAlLambda,PerlmutterEtAlLambda}, which undergo an accelerated expansion. The decay of classical fields propagating on dS spacetimes and its generalizations containing black holes---such as Schwarzschild--de~Sitter (SdS), Kerr--de~Sitter (KdS), Reissner--Nordstr\"om--de~Sitter (RNdS), and Kerr--Newman--de~Sitter (KNdS) spacetimes---has been studied in detail via numerical evolutions \cite{BradyChambersKrivanLagunaCosmConst,BradyChambersLaarakeersPoissonSdSFalloff,LunaZilhaoCardosoCostaNatarioSCCNonlinear}, the computation of quasinormal modes (QNMs) \cite{MellorMossStability,BradyMossMyersCosmicCensorship,AbdallaCastelloBrancoLimaSantosdSCFT,CardosoLemosSdSQNM,MyungKimdS,DuWangSuQNMdS,NatarioSchiappaQNM,ChoudhuryPadmanabhanQNMforSdS,LopezQNMdS2006d34,LopezQNMdS2006D,LopezQNMDiracD,BertiCardosoStarinetsQNM,LopezQNMdS2012,AnsorgMacedoHyp,MacedoJaramilloAnsorgQNMHypRN,CardosoCostaDestounisHintzJansenSCC,MacedoKerrQNM,BizonChmajMachHyperboloidal}, and mathematical investigations \cite{FriedrichStability,SaBarretoZworskiResonances,BonyHaefnerDecay,RingstromEinsteinScalarStability,YagdjianGalstianFundamentalSolKG,VasyWaveOndS,DyatlovAsymptoticDistribution,DyatlovZworskiTrapping,YagdjianGlobalScalarField,SchlueCosmological,SchlueWeylDecay,HintzVasyKdSStability,HintzKNdSStability}.

On dS, which has a single temporal scale given by the cosmological horizon, QNMs capture the temporal dependence at late times of linear and weakly nonlinear fields $\Phi$ \cite{VasyMicroKerrdS,HintzVasySemilinear}, in that
\begin{equation}
\label{EqIntro}
  \Phi(t,x) \sim \sum e^{-i\omega_j t} c_j u_j(x),\quad t\to\infty,
\end{equation}
where the $u_j$ are normalized mode solutions, and the expansion coefficients $c_j$ are determined by the initial conditions of $\Phi$. On black hole spacetimes such as SdS, QNMs capture also the ringdown phase \cite{BertiCardosoStarinetsQNM} on the much smaller temporal scale given by the mass of the black hole, as recently demonstrated also experimentally via gravitational wave measurements \cite{LIGOBlackHoleMerger,IsiGieslerFarrScheelTeukolskyNoHair}. There has also been interest in QNMs of asymptotically dS and anti--de~Sitter (AdS) black holes owing to the dS/CFT \cite{StromingerdSCFT,AbdallaCastelloBrancoLimaSantosdSCFT} and AdS/CFT correspondences \cite{MaldacenaAdSCFT,HorowitzHubenyAdSQNM}.

Our objective is to clarify the existence and relevance of QNMs for massive scalar fields, i.e.\ solutions of the Klein--Gordon equation, on (the static model of) dS, and conclusively explain the discrepancies between contradictory results reported in the literature. Early calculations \cite{BradyChambersLaarakeersPoissonSdSFalloff,ChoudhuryPadmanabhanQNMforSdS} find the correct QNMs but discard some of them due to observed pole cancellations in explicit expressions for the frequency space Green's function and by comparison with numerical wave evolutions. The pole cancellations are ignored (hence, all QNMs are kept) in \cite{AbdallaCastelloBrancoLimaSantosdSCFT}, while \cite{MyungKimdS} finds no QNMs at all, and \cite{DuWangSuQNMdS} finds QNMs only under a condition on the scalar field mass. The article \cite{NatarioSchiappaQNM} asserts the existence of QNMs only for dS with odd spacetime dimension $D$. The series of papers \cite{LopezQNMdS2006d34,LopezQNMdS2006D,LopezQNMDiracD,LopezQNMdS2012} computes QNMs for a variety of classical hyperbolic equations on dS.

We will demonstrate that QNMs in fact exist for massive scalar fields on dS in all spacetime dimensions $D\geq 2$ and for all scalar field masses (real or complex). The pole cancellations and matching wave evolutions are shown to be due to the explicit or implicit assumption that the initial data of the field are equal to $0$ near the dS horizon. Without this restrictive assumption, no QNMs may be discarded; every frequency for which there exists a purely outgoing mode solution of the field equation \emph{is} a QNM and contributes to the QNM expansion~\eqref{EqIntro}.

The key is to understand the dependence of the coefficients $c_j$ in~\eqref{EqIntro} on the initial conditions of the field: $c_j$ can typically be computed as the inner product of an expression formed from the initial data with a purely \emph{incoming} mode solution $v_j(x)$ with frequency $\bar\omega$. In certain exceptional situations however, for example for massless scalar fields on dS in $D=4$ dimensions and for all nonzero QNMs, no purely incoming mode solution exists. Instead, the role of $v_j$ is now played by a \emph{distribution} which is supported on the dS horizon, i.e.\ $v_j$ is a sum of differentiated $\delta$-distributions. In these exceptional situations, it is the behavior of the initial data in an arbitrarily small neighborhood of the horizon which determines $c_j$. We give a characterization of $v_j$ which is valid in all cases, and call such generalized purely incoming solutions \emph{dual resonant states}.\footnote{This term has previously been used in the mathematics literature \cite{HintzVasyKdsFormResonances,HintzVasyKdSStability}; another common term is \emph{co-resonant state} \cite{GuillarmouHilgertWeichDensities}.} This characterization requires working in a hyperboloidal slicing of dS \cite{AnsorgMacedoHyp,MacedoJaramilloAnsorgQNMHypRN,BizonChmajMachHyperboloidal}; the failure of time evolution to be unitary in this slicing is the reason for the asymmetry between the definitions of mode solutions $u_j$ and dual resonant states $v_j$. The notion of dual resonant states thus allows one to discern what triggers the appearance of any particular QNM frequency in the expansion~\eqref{EqIntro}.

On dS specifically, we compute a few dual resonant states for a variety of dS dimensions and scalar field masses. We also present a simple method to determine QNMs and mode solutions for wave type equations on pure dS, based on an efficient description of the asymptotic behavior of waves on the conformal compactification of dS in terms of data on the conformal boundary \cite{VasyWaveOndS,HintzVasyKdSStability,HintzXieSdS} and the solution of a certain characteristic polynomial. To the authors' knowledge, this method appears here for the first time. We illustrate our results with the numerical evolution of massive scalar waves in hyperboloidal slicing \cite{BizonRostworowskiZenginogluYM}.

%%%%%%%%%%%%%%%%%%%%%%%%%%%%%%%%%%%%%%%%%%%%%%%%%%
\section{LATE-TIME BEHAVIOR OF MASSIVE SCALAR FIELDS}

The line element of the static model of $D$-dimensional dS is given by
\begin{equation}
\label{EqdSStatic}
  \dd s^2 = -F(r)\,\dd t^2 + \frac{\dd r^2}{F(r)} + r^2\dd\Omega^2,\quad
  F(r)=1-\frac{r^2}{L^2},
\end{equation}
where $\dd\Omega^2$ is the line element of the $(D-2)$-dimensional unit sphere, and the dS radius $L$ is related to $\Lambda$ and the surface gravity $\kappa$ of the dS horizon via
\begin{equation}
  \Lambda = \frac{(D-1)(D-2)}{2 L^2},\quad
  \kappa = L^{-1}.
\end{equation}
The level sets of the time function
\begin{equation}
  t_* = t + \frac{1}{2\kappa}\log F
\end{equation}
are hyperboloidal, i.e.\ transversal to the future cosmological horizon, and the dS line element
\begin{equation}
  \dd s^2 = -F(r)\,\dd t_*^2 - \frac{2 r}{L}\dd t_*\dd r + \dd r^2 + r^2\dd\Omega^2
\end{equation}
is regular across the horizon $r=L$. Finally, regarding $r,\Omega$ as the polar coordinates of $x=r\Omega$ and setting
\begin{equation}
\label{EqtauX}
  \tau = e^{-\kappa t_*},\quad
  X = \frac{x\tau}{L},
\end{equation}
we obtain the expression
\begin{equation}
\label{EqConf}
  \dd s^2 = L^2\frac{-\dd\tau^2 + \dd X^2}{\tau^2},
\end{equation}
or equivalently the more familiar FLRW form of the dS metric, $\dd s^2=-\dd t_*^2 + e^{-2\kappa t_*}\dd(L X)^2$.

%%%%%%%%%%%%%%%%%%%%%%%%%%%%%%%%%%%%%%%%%%%%%%%%%%
\subsection{DATA ON THE CONFORMAL BOUNDARY}

In terms of \eqref{EqConf}, the equation of motion $(\Box-m^2)\Phi=0$ for the scalar field $\Phi=\Phi(\tau,X)$ of mass $m$ reads
\begin{equation}
\label{EqKG}
  L^{-2}\bigl(-(\tau\pa_\tau)^2 + (D-1)\tau\pa_\tau + \tau^2\Delta_X\bigr)\Phi - m^2\Phi = 0,
\end{equation}
where $\Delta_X=\sum\pa_{X^i}^2$ is the spatial Laplace operator. We are interested in the behavior of $\Phi$ as $\tau\to 0$ (i.e.\ $t_*\to\infty$), with initial data
\begin{equation}
\label{EqKGID}
  (\Phi,\pa_{t_*}\Phi)|_{t_*=0} = (\Phi,-\tau\pa_\tau\Phi)|_{\tau=1} = (\Phi_0,\Phi_1)
\end{equation}
which we assume to be smooth functions of $X$. Dropping the term $\tau^2\Delta_X$ in~\eqref{EqKG}, the characteristic exponents of the resulting ODE in $\tau$ at $\tau=0$ are the roots
\begin{equation}
  \lambda_\pm(m) = \frac{D-1}{2} \pm \sqrt{\frac{(D-1)^2}{4}-L^2 m^2}
\end{equation}
of the quadratic polynomial
\begin{equation}
  p(\lambda) = \lambda^2 - (D-1)\lambda + L^2 m^2.
\end{equation}
Let $\Delta\lambda=\lambda_+(m)-\lambda_-(m)=\sqrt{(D-1)^2-4 L^2 m^2}$.

\textit{The generic case $\Delta\lambda\notin 2\N_0$.---} It can be shown \cite{VasyWaveOndS,HintzXieSdS} that in this case
\begin{equation}
\label{EqPhi}
  \Phi(\tau,X) = \sum_\pm \tau^{\lambda_\pm(m)}u_\pm(\tau,X),
\end{equation}
where $u_\pm(\tau,X)$ are smooth functions of $(\tau,X)\in[0,1]\times\R^D$ whose Taylor expansion at $\tau=0$ only contains even powers of $\tau$. Moreover, there is a one-to-one correspondence
\begin{equation}
\label{EqCorr}
  (\Phi_0,\Phi_1) \leftrightarrow (u_+^{(0)},u_-^{(0)}) := (u_+,u_-)|_{\tau=0}
\end{equation}
between the initial data $(\Phi_0,\Phi_1)$ of $\Phi$ and the asymptotic data $(u_+^{(0)},u_-^{(0)})$.

We sketch the construction of a solution $\Phi$ of~\eqref{EqKG} given asymptotic data $(u_+^{(0)},u_-^{(0)})$: we make the ansatz
\begin{equation}
\label{EqAnsatz}
  u_\pm(\tau,X) = \sum_{j=0}^\infty \tau^{2 j}u_\pm^{(j)}(X)
\end{equation}
in~\eqref{EqPhi} and plug this into~\eqref{EqKG}. This gives a recursion relation for the functions $u_\pm^{(j)}$, $j\geq 1$, with unique solution
\begin{equation}
\label{EqAnsatzSol}
  u_\pm^{(j)} = q_\pm^{(j)}\Delta_X^j u_\pm^{(0)},\qquad q_\pm^{(j)}=\prod_{k=1}^j p(\lambda_\pm(m)+2 k)^{-1}.
\end{equation}
When $u_\pm^{(0)}$ are analytic functions of $X$, then~\eqref{EqAnsatz} converges near $\tau=0$; see Appendix~\ref{AppConv}. When $u_\pm^{(0)}$ are merely smooth, then, as shown in \cite{HintzXieSdS}, one can still find $u_\pm$ with Taylor coefficients at $\tau=0$ given by the $u_\pm^{(j)}$ in~\eqref{EqAnsatz} so that $\Phi$ solves~\eqref{EqKG} exactly.

\textit{The exceptional case $\Delta\lambda\in 2\N_0$.---} This occurs when $m^2=\frac{(D-1)^2-4 n^2}{4 L^2}$ for some $n\in\N_0$, thus in any fixed dimension $D$ only for a finite number of \emph{real} scalar field masses $m$. The asymptotics of the field are now
\begin{equation}
\label{EqExceptional}
  \Phi(\tau,X) = \tau^{\lambda_-(m)}u_-(\tau,X) + \tau^{\lambda_+(m)}(\log\tau)u_+(\tau,X),
\end{equation}
and the one-to-one correspondence~\eqref{EqCorr} still holds: the full Taylor series of $u_\pm$ at $\tau=0$ are determined by $u_\pm^{(0)}$. We omit the explicit formulae.

%%%%%%%%%%%%%%%%%%%%%%%%%%%%%%%%%%%%%%%%%%%%%%%%%%
\subsection{QNMS AND MODE SOLUTIONS}

Quasinormal modes $\omega$ are typically defined in terms of static coordinates~\eqref{EqdSStatic} as those complex numbers $\omega\in\C$ for which there exists a purely outgoing solution
\begin{equation}
\label{EqQNMOut}
  e^{-i\omega t}u(r,\Omega),\qquad
  u(r\to L) \sim e^{i\omega r_*},
\end{equation}
of the Klein--Gordon equation $(\Box-m^2)(e^{-i\omega t}u)=0$; here we introduced the tortoise coordinate $\dd r_*=\dd r/F$. Near $r=0$, $u$ is required to be bounded, which automatically implies the smoothness of $u$ as a function of $x=r\Omega\in\R^D$ near $x=0$ \cite{HintzXieSdS}. Now $t_*-(t-r_*)=L\log(1+\frac{r}{L})$ is analytic in $r>0$ across $r=L$, hence outgoing solutions are of the form
\begin{equation}
\label{EqQNMSmooth}
  e^{-i\omega t_*}u,\qquad
  u\text{ smooth near $x=r\Omega=0$ and $|x|=L$}.
\end{equation}
Solutions\footnote{For analytic spacetimes and time functions $t_*$, $u$ is automatically analytic \cite{GalkowskiZworskiHypo,PetersenVasyAnalytic}.} of the field equation of the form~\eqref{EqQNMSmooth} are smooth across the future cosmological horizon of dS, hence QNMs are indeed those $\omega$ for which a nontrivial solution of this form exists \cite{VasyMicroKerrdS,WarnickQNMs,BizonChmajMachHyperboloidal,JansenMathematica,HintzXieSdS}. We remark that smoothness prohibits incoming asymptotics $e^{-i\omega t}e^{-i\omega r_*}\sim e^{-i\omega t_*}e^{-2 i\omega r_*}\sim e^{-i\omega t_*}|L-r|^{i\omega/2\kappa}$ as $r\to L$ for all but an exceptional set of values of $\omega$.

For future use, we record that the PDE solved by $u$ in~\eqref{EqQNMSmooth} is
\begin{equation}
\label{EqQNMEq}
\begin{split}
  \Bigl(r^{-D+2}\pa_r r^{D-2}F\pa_r &+ i\kappa\omega(2 r\pa_r+D-1) \\
  &\qquad + r^{-2}\Delta_\Omega + \omega^2 - m^2\Bigr)u = 0,
\end{split}
\end{equation}
where $\Delta_\Omega$ is the Laplacian on the $(D-2)$-sphere.

Denoting by $\omega_j$ the QNMs and by $u_j=u_j(x)$ the corresponding mode solutions (normalized arbitrarily), the quasinormal mode expansion of the massive scalar field is
\begin{equation}
\label{EqQNMExp1}
  \Phi(t_*,x) \sim \sum c_j e^{-i\omega_j t_*}u_j(x),\qquad t_*\to\infty,
\end{equation}
for certain complex coefficients $c_j$ depending on the initial conditions of the field.\footnote{The notation `$\sim$' means that for any $C$, the difference between $\Phi$ and the sum over all $\omega_j$ with $\Im\omega_j\geq-C$ is bounded by $C' e^{-C t_*}$.} \footnote{For QNMs with higher multiplicity---ignored here---there are additional terms with time dependence $e^{-i\omega_j t_*}t_*^k$.} On the other hand, note that $t_*\to\infty$ while keeping $x$ bounded implies $\tau,X\to 0$. If we Taylor expand~\eqref{EqPhi} at $(\tau,X)=(0,0)$ and undo the coordinate change~\eqref{EqtauX}, we get the expansion
\begin{equation}
\label{EqQNMExp2}
  \Phi(t_*,x) \sim \sum_\pm\sum_{n=0}^\infty e^{-\kappa(\lambda_\pm(m)+n)t_*} \sum_{2 j+|\alpha|=n} c_\pm^{(j,\alpha)}\Bigl(\frac{x}{L}\Bigr)^\alpha,
\end{equation}
where $c_\pm^{(j,\alpha)}=\frac{1}{\alpha!}q_\pm^{(j)}\pa_X^\alpha \Delta_X^j u_\pm^{(0)}(0)$. By comparing this with~\eqref{EqQNMExp1}, we can read off the QNMs of the massive scalar field to be those $\omega\in\C$ for which $e^{-i\omega t_*}=e^{-\kappa(\lambda_\pm(m)+n)t_*}$ for some $n=0,1,2,\ldots$; thus, the QNMs are
\begin{equation}
\label{EqdSQNM}
  -i\kappa(\lambda_-(m)+n),\ -i\kappa(\lambda_+(m)+n),\quad  n=0,1,2,\ldots
\end{equation}

Moreover, we can directly read off all mode solutions corresponding to any one of these QNMs, i.e.\ all functions $u=u(x)$ so that $e^{-\kappa(\lambda_\pm(m)+n)t_*}u(x)$ (with the choice of sign and $n=0,1,2,\ldots$ fixed) solves the equation~\eqref{EqKG} for the scalar field. Indeed, the freedom in the coefficients $c_\pm^{(j,\alpha)}$, with $n=2 j+|\alpha|$ fixed (thus considering the QNM $-i\kappa(\lambda_\pm(m)+n)$), is fully accounted for by the freedom of specifying all derivatives $\pa_X^\beta u_\pm^{(0)}(0)$ of order $|\beta|=n$ (there are $\bigl(\genfrac{}{}{0pt}{}{n+D-2}{D-2}\bigr)$ of them). As one varies the complex numbers $c_\pm^{(j,\alpha)}$ freely, the innermost sum in~\eqref{EqQNMExp2} computes all elements of the corresponding $\bigl(\genfrac{}{}{0pt}{}{n+D-2}{D-2}\bigr)$-dimensional space of mode solutions.\footnote{In the exceptional case~\eqref{EqExceptional}, the QNMs are still given by~\eqref{EqdSQNM}, and the values which are accounted for twice have multiplicity two.} We stress that one can choose the asymptotic data $u_\pm^{(0)}$ so that the innermost sum in~\eqref{EqQNMExp2} is a nonzero function for any desired value of $n$, i.e.\ for all of the QNMs in~\eqref{EqdSQNM} a nontrivial mode solution indeed exists, and thus all QNMs contribute to the long-time dynamics~\eqref{EqQNMExp1} of the field.

In the literature, QNMs on spacetimes with spherical symmetry, such as dS, are typically computed by separating angular and radial parts of putative mode solutions $u(x)$ via expansion into spherical harmonics, and by studying a radial ODE for each angular momentum $l$. In this manner, one obtains a set of QNMs for each $l=0,1,2,\ldots$, consisting of those $\omega\in\C$ for which there exists a separated mode solution with angular momentum $l$. In order to relate this to our result~\eqref{EqQNMExp2}, observe that in the decomposition of a summand $x^\alpha$ (appearing in~\eqref{EqQNMExp2}) into a sum of products of spherical harmonics and functions of $r=|x|$, degree $l$ spherical harmonics appear if and only if $|\alpha|-l\in 2\N_0$.\footnote{As an example, in $\R^3$ with coordinates $x=(x_1,x_2,x_3)$, one has $x_3^3=r^3\sqrt{12\pi/25}Y_{1 0}(\theta,\phi)+r^3\sqrt{16\pi/175}Y_{3 0}(\theta,\phi)$; note that only $Y_{l m}$ with $l=1,3$ appear.} That is, for a QNM $\omega=-i\kappa(\lambda_\pm(m)+n)$, the mode solution $\sum_{2 j+|\alpha|=n}c_\pm^{(j,\alpha)}(x/L)^\alpha$, expanded into spherical harmonics, has a nontrivial piece with angular momentum $l$ only if there is a nonzero summand with $|\alpha|-l=2 \tilde n$ for some $\tilde n\in\N_0$, so
\begin{equation}
  n=2 j+|\alpha|=2 n'+l
\end{equation}
where $n'=j+\tilde n\in\N_0$. Thus, restricting to mode solutions of the separated form $u(r,\Omega)=Y(\Omega)u_0(r)$ where $Y$ is a degree $l$ spherical harmonic, the QNM spectrum is equal to
\begin{equation}
  -i\kappa(\lambda_-(m)+l+2 n'),\ -i\kappa(\lambda_+(m)+l+2 n'),
\end{equation}
with $n'=0,1,2,\ldots$. (The fact that every QNM is necessarily of this form is well-known \cite{BradyChambersLaarakeersPoissonSdSFalloff,LopezQNMdS2006D}.)

%%%%%%%%%%%%%%%%%%%%%%%%%%%%%%%%%%%%%%%%%%%%%%%%%%%%%%%%%%%%%%%%%%%%%%
\section{DUAL RESONANT STATES}

So far, we have given a new perspective on the well-known result (see e.g.\ \cite[\S IV.C]{BradyChambersLaarakeersPoissonSdSFalloff}) that for each frequency~\eqref{EqdSQNM} there exists an outgoing mode solution of the Klein--Gordon equation on pure dS. On the other hand, the authors in \cite[\S IV.D]{BradyChambersLaarakeersPoissonSdSFalloff} observe a pole cancellation in the Wronskian of outgoing and incoming solutions at frequencies $\omega$ for which
\begin{equation}
\label{EqDeltaQNM}
  \omega=-i n\kappa,\qquad n=1,2,3,\ldots
\end{equation}
and consequently discard such $\omega$ from the set of QNMs. (Similar arguments are presented for other classes of perturbations and in general dimension in \cite[\S4.3]{NatarioSchiappaQNM}: again, certain QNMs are discarded despite the existence of outgoing mode solutions.) Moreover, wave evolutions performed in \cite[\S IV.D]{BradyChambersLaarakeersPoissonSdSFalloff}, with initial data supported inside the dS horizon, show that the late time asymptotics of mass $m$ scalar fields change qualitatively---seemingly consistent with discarding the values~\eqref{EqDeltaQNM} in the QNM expansion~\eqref{EqQNMExp1}---when $m$ approaches a value $m_0$ for which all nonzero QNMs listed in~\eqref{EqdSQNM} satisfy~\eqref{EqDeltaQNM}. (This happens for $m_0=0$ and for the conformal mass $m_0=\tfrac12\sqrt{D(D-2)}$.)

We shall reconcile our late time asymptotics~\eqref{EqQNMExp2} (leading to the \emph{full} set of QNMs~\eqref{EqdSQNM}) with the wave evolution in \cite{BradyChambersLaarakeersPoissonSdSFalloff} by analyzing the dependence of the coefficients $c_j$ in~\eqref{EqQNMExp1} on the initial data~\eqref{EqKGID} of $\Phi$, ultimately showing in~\S\ref{SWave} that the qualitative change observed in \cite[\S IV.D]{BradyChambersLaarakeersPoissonSdSFalloff} is \emph{not} present when the initial data are allowed to be nonzero near the dS horizon. In~\S\ref{SGreen}, we shall also connect our analysis to the Green's function computations presented in \cite{BradyChambersLaarakeersPoissonSdSFalloff,ChoudhuryPadmanabhanQNMforSdS,NatarioSchiappaQNM}.

The following arguments are very general and apply not only to the Klein--Gordon equation on dS, but generalize in a straightforward manner to any (not necessarily scalar) wave equation on any stationary spacetime foliated by level sets (of a time function $t_*$) which are transversal to all future horizons; this in particular includes KNdS spacetimes (and its special cases SdS, KdS, RNdS), and with some modifications also asymptotically flat ($\Lambda=0$) and anti--de~Sitter ($\Lambda<0$) black hole spacetimes. We study the forced equation
\begin{equation}
\label{EqForced}
  (\Box-m^2)\Phi = f,\quad \Phi|_{t_*<0}=0,
\end{equation}
where the forcing $f=f(t_*,x)$ satisfies $f=0$ for $t_*<0$ and $t_*\gg 1$. We remark that this covers also initial value problems, in the following manner: denoting, for clarity, by $\Phi_{\rm IVP}=\Phi_{\rm IVP}(t_*,X)$ the solution of~\eqref{EqKG}--\eqref{EqKGID}, let $\Phi(t_*,X)=\Theta(t_*)\Phi_{\rm IVP}(t_*,X)$ where $\Theta$ is the Heaviside function. Then $(\Box-m^2)\Phi_{\rm IVP}=0$ allows us to write
\begin{equation}
\label{EqComm}
\begin{split}
  (\Box-m^2)\Phi &= (\Box-m^2)(\Theta(t_*)\Phi_{\rm IVP}) \\
    &= [\Box-m^2,\Theta(t_*)]\Phi_{\rm IVP} =: f,
\end{split}
\end{equation}
where $[A,B]=A B-B A$ is the commutator. But since $\Box-m^2$ is a second order differential operator, we have $[\Box-m^2,\Theta(t_*)]=\delta(t_*)A_1(X,D_X)+\delta'(t_*)A_0(X)$, where $A_0$ is a smooth function and $A_1$ is a first order differential operator. In particular, for the calculation of $f$ in~\eqref{EqComm} one may replace $\Phi_{\rm IVP}$ by its linearization at $t_*=0$ which in terms of~\eqref{EqKGID} is $\Phi_0+t_*\Phi_1$. Thus, the solution of the initial value problem~\eqref{EqKG}--\eqref{EqKGID} in $t_*>0$ is equal to the solution of the forced equation~\eqref{EqForced} for $f=[\Box-m^2,\Theta(t_*)](\Phi_0+t_*\Phi_1)$.

We now return to the equation~\eqref{EqForced} for general $f$. Assuming that the $j$-th QNM $\omega_j$ is simple, the map assigning to $f$ the coefficient $c_j$ in the QNM expansion~\eqref{EqQNMExp1} is linear, hence given by the formula
\begin{equation}
\label{EqDualcj}
  c_j[f] = \iint f(t_*,x) \overline{\Psi_j(t_*,x)}\,\dd t_*\,\dd^{D-1}x
\end{equation}
for some (necessarily nonzero) distribution $\Psi_j$. We shall deduce the key properties of $\Psi_j$ by plugging special choices of $f$ into~\eqref{EqForced}--\eqref{EqDualcj}.

First, for any $f$ so that $f\equiv 0$ inside the horizon (i.e.\ for $r=|x|\leq L$ in the dS case), we have $\Phi\equiv 0$ inside the horizon by finite speed of propagation for~\eqref{EqForced}; therefore,
\begin{equation}
\label{EqDualSupp}
  \Psi_j(t_*,x)=0\ \text{outside the horizon}\ \text{(i.e.\ for $r>L$ on dS)}.
\end{equation}
Next, if $f=(\Box-m^2)g$ for a function $g$ with $g=0$ for $t_*<0$ and  $t_*\gg 1$, then the solution of~\eqref{EqForced} is given by $\Phi=g$ and thus vanishes for large $t_*$, hence $c_j[(\Box-m^2)g]=0$. Plugging this into~\eqref{EqDualcj} and integrating by parts, we find that $\Psi_j$ must solve the adjoint equation\footnote{Thus $\Psi_j$ can be interpreted as a massive scalar field with mass $\bar m$ ($=m$ for real masses) which propagates backwards in time.}
\begin{equation}
\label{EqDualAdjEq}
  (\Box-m^2)^*\Psi_j = (\Box-\bar m^2)\Psi_j = 0.
\end{equation}
Finally, if $\Phi$ solves~\eqref{EqForced}, then $\Psi=(\pa_{t_*}+i\omega_j)\Phi$ solves $(\Box-m^2)\Psi=(\pa_{t_*}+i\omega_j)f$; but since $(\pa_{t_*}+i\omega_j)e^{-i\omega_j t_*}=0$, the $j$-th coefficient in the expansion of $\Psi$ is zero, so $c_j[(\pa_{t_*}+i\omega_j)f]=0$. Integrating by parts in~\eqref{EqDualcj} implies that
\begin{equation}
\label{EqDualMode}
  (-\pa_{t_*}-i\bar\omega_j)\Psi_j = 0\ \implies\ \Psi_j(t_*,x)=e^{-i\bar\omega_j t_*}v_j(x).
\end{equation}
A crucial point is that $v_j$ may vanish also inside the horizon, i.e.\ in $r<L$ on dS, in which case it must be a sum of differentiated $\delta$-distributions \emph{supported at the horizon}. This happens e.g.\ for $D=4$, $m=0$ or $m=\sqrt{2}$, with important consequences for wave evolution and the pole structure of the Green's function, as discussed later.

Summarizing~\eqref{EqDualSupp}--\eqref{EqDualMode}, we arrive at the following definition: a \emph{dual resonant state} at frequency $\omega$ is a distribution\footnote{One can show that $v$ is necessarily smooth where $\pa_{t_*}$ is timelike, hence on dS $v$ can be singular only at $r=L$.} $v=v(x)$ such that
\begin{equation}
\label{EqDualDef}
 \boxed{
  \begin{gathered}
    (\Box-\bar m^2)(e^{-i\bar\omega t_*}v)=0, \\
    v=0\ \text{outside the horizon}\ \text{(i.e.\ for $r>L$ on dS).}
  \end{gathered}
 }
\end{equation}
We have moreover proved the existence of a dual resonant state for every QNM frequency.

One can show \cite{VasyMicroKerrdS} that the space of solutions of~\eqref{EqDualDef} for fixed $\omega$ has the same dimension as the space of mode solutions~\eqref{EqQNMSmooth}. On spherically symmetric spacetimes such as dS or RNdS, these spaces are typically $1$-dimensional upon restricting to modes or dual resonant states which are of the separated form $w(r)Y(\Omega)$ where $Y(\Omega)$ is a fixed spherical harmonic.

Assume that for the QNM $\omega_j$ there indeed exist a unique mode solution $u_j$ and dual resonant state $v_j$ (up to scalar multiples, and restricting to fixed angular dependence if necessary). We shall determine the normalization constant $a_j$ so that
\begin{equation}
\label{EqDualNorm}
  \Psi_j=a_j e^{-i\bar\omega_j t_*}v_j
\end{equation}
computes $c_j[f]$ in~\eqref{EqDualcj}. To this end, note that the function $\Phi=\Theta(t_*)e^{-i\omega_j t_*}u_j$ solves~\eqref{EqForced} with forcing $f=[\Box-m^2,\Theta(t_*)](e^{-i\omega_j t_*}u_j)$, as follows by the same calculation as~\eqref{EqComm}. Since the QNM expansion of $\Phi$ has $c_j=c_j[f]=1$, the constant $a_j$ in~\eqref{EqDualNorm} satisfies
\begin{equation}
\label{EqDualNormC}
  a_j\iint [\Box{-}m^2,\Theta(t_*)](e^{-i\omega_j t_*}u_j) \overline{e^{-i\bar\omega_j t_*}v_j}\,\dd t_*\,\dd^{D-1}x = 1.
\end{equation}
Defining the spectral family
\begin{equation}
\label{EqSpecFam}
  P(\omega) := e^{i\omega t_*}(\Box-m^2)e^{-i\omega t_*}
\end{equation}
(which on dS is equal to the operator in parentheses in~\eqref{EqQNMEq}), we thus obtain from~\eqref{EqDualcj}, \eqref{EqDualNorm}, \eqref{EqDualNormC} after a brief calculation the formula
\begin{equation}
\label{EqDualcjFinal}
  c_j[f] = i\frac{\int \hat f(\omega_j,x)\overline{v_j(x)}\,\dd^{D-1}x}{\int (\pa_\omega P(\omega_j)u_j)\overline{v_j}\,\dd^{D-1}x},
\end{equation}
where $\hat f(\omega,x)=\int e^{i\omega t_*}f(t_*,x)\,\dd t_*$ is the Fourier transform in time. This is directly related to the pole structure of the Green's function $G(\omega;x,x')$ of $P(\omega)$, which near $\omega_j$ takes the form
\begin{equation}
\label{EqGreensPole}
  G(\omega;x,x') = \frac{u_j(x)\overline{v_j(x')}}{(\omega-\omega_j)\int (\pa_\omega P(\omega_j)u_j)\overline{v_j}\,\dd^{D-1}x} + {\rm{hol.}}
\end{equation}
It is important here that $x,x'$ are not restricted to lie inside the horizon.

%%%%%%%%%%%%%%%%%%%%%%%%%%%%%%%%%%%%%%%%%%%%%%%%%%
\subsection{CONNECTION WITH PURELY INCOMING MODES} 
%%%%%%%%%%%%%%%%%%%%%%%%%%%%%%%%%%%%%%%%%%%%%%%%%%

Taking the adjoint of~\eqref{EqQNMEq} (or directly from~\eqref{EqDualDef}), a dual resonant state $v$ on dS at frequency $\omega$ and with angular dependence given by a fixed degree $l$ spherical harmonic solves the equation
\begin{equation}
\label{EqDualEq}
\begin{split}
  &P(\omega)^*v=\Bigl(r^{-D+2}\pa_r r^{D-2}F\pa_r + i\kappa\bar\omega(2 r\pa_r+D-1) \\
  &\hspace{6em} - r^{-2}l(l+D-3) + \bar\omega^2 - \bar m^2\Bigr)v = 0.
\end{split}
\end{equation}
Multiplying this equation by $F$ and using that $\pa_r\sim -2\kappa\pa_F$ near $r=L=\kappa^{-1}$ (where $F\approx 0$), it reads
\begin{equation}
  \bigl(4\kappa^2(F\pa_F)^2 - 4 i\bar\omega\kappa F\pa_F\bigr)v \approx 0\ \text{for}\ F\approx 0.
\end{equation}
This has solutions $F^z$ with homogeneity $z=0$ and $z=i\bar\omega/\kappa$. One can show \cite{VasyMicroKerrdS} that the dual resonant state must have the latter homogeneity (i.e.\ precisely the behavior disallowed for outgoing modes), so
\begin{equation}
\label{EqIncoming}
  e^{-i\bar\omega t_*}v = e^{-i\bar\omega t}v^s,\qquad v^s(r\to L)\sim e^{-i\bar\omega r_*}
\end{equation}
upon restricting to the static patch of dS. Thus, $v^s$ is a purely incoming mode with frequency $\bar\omega$; or $v^s\equiv 0$ if $v$ is supported on $r=L$. The latter situation can only happen in special circumstances: a distribution supported on $F=0$ has as its leading order term $\delta^{(n-1)}(F)$ for some $n=1,2,3,\ldots$, which is homogeneous of degree $-n$. Therefore, $v$ can be supported on the dS horizon if and only if $i\bar\omega/\kappa=-n$, which is equivalent to condition~\eqref{EqDeltaQNM}. If $\omega$ is \emph{not} of this form, then a dual resonant state at frequency $\omega$ is the same (upon extension by $0$ beyond the horizon) as a purely incoming mode as in~\eqref{EqIncoming}.

We can gain further insight by re-writing the equation $P(\omega)u=0$ satisfied by a mode solution $u=u(r,\Omega)$ in static coordinates~\eqref{EqdSStatic}, so $P^s_m(\omega)u^s=0$ where $u^s=F^{-i\omega/2\kappa}u$ and
\begin{equation}
\label{EqSpectralFy}
  P^s_m(\omega) = r^{-D+2}\pa_r r^{D-2}F\pa_r + r^{-2}\Delta_\Omega + F^{-1}\omega^2-m^2.
\end{equation}
Since only the square of $\omega$ appears in the coefficients of $P^s_m(\omega)$, taking complex conjugates gives
\begin{equation}
  0 = \overline{P^s_m(\omega)u^s} = P^s_{\bar m}(\bar\omega)\overline{u^s} = P^s_{\bar m}(-\bar\omega)\overline{u^s}.
\end{equation}
Therefore, $(\Box-\bar m^2)(e^{i\bar\omega t}\overline{u^s})=0$, and hence
\begin{equation}
  (\Box-\bar m^2)(e^{i\bar\omega t_*} F^{i\bar\omega/\kappa}\bar u)=0.
\end{equation}
But $F^{i\bar\omega/\kappa}\bar u$ is purely incoming at the horizon, so taking into account the vanishing requirement for dual states, one is tempted to conclude that\footnote{We use the notation $x_+=\max(0,x)$.}
\begin{equation}
\label{EqDualGuess}
  v = F_+^{i\bar\omega/\kappa}\bar u = (1-r^2/L^2)_+^{i\bar\omega/\kappa}\bar u
\end{equation}
is a dual state. And indeed, as long as $z:=i\bar\omega/\kappa$ is not a negative integer (i.e.\ $\omega$ is not of the form~\eqref{EqDeltaQNM}), \emph{this is correct}.\footnote{Indeed, by construction, $v$ solves~\eqref{EqDualEq} in $r<L$ and $r>L$, hence $P(\omega)^*v$ must be supported at $r=L$, so it is a sum of differentiated $\delta$-distributions. But $\delta^{(n-1)}(F)$ is homogeneous of degree $-n\in-\N$ at $F=0$, whereas $v$ is homogeneous of degree $i\bar\omega/\kappa\notin-\N$ to leading order at $F=0$. Therefore, necessarily $P(\omega)^*v=0$.} The delicate point is that the distribution $F_+^z$ depends only \emph{meromorphically} on $z$ and has simple poles at $z\in-\N$. Its regularization $\chi_+^z(F)=\Gamma(z+1)^{-1}F_+^z$ on the other hand is well-defined (see \cite[\S III.3.2]{HormanderAnalysisPDE1}) and satisfies $\chi_+^{-n}(F)=\delta^{(n-1)}(F)$. Thus, in the exceptional case~\eqref{EqDeltaQNM}, one needs to replace~\eqref{EqDualGuess} by
\begin{equation}
\label{EqDualGuess2}
  v = \delta^{(n-1)}(F)\bar u,\qquad n=-i\bar\omega/\kappa\in\N.
\end{equation}

%%%%%%%%%%%%%%%%%%%%%%%%%%%%%%%%%%%%%%%%%%%%%%%%%%
\subsection{EXPLICIT EXAMPLES}
\label{SExpl}
%%%%%%%%%%%%%%%%%%%%%%%%%%%%%%%%%%%%%%%%%%%%%%%%%%

For concreteness, we fix the scale of the cosmological horizon to be
\begin{equation}
  L=1.
\end{equation}
As an independent verification of formula~\eqref{EqDualGuess2}, we may find those dual states (corresponding to QNMs satisfying~\eqref{EqDeltaQNM}) which are sums of differentiated $\delta$-distributions as follows. We make the ansatz
\begin{equation}
\label{EqSumDelta}
  v(r)=a_0\delta(r-1)+a_1\delta'(r-1)+\cdots+a_n\delta^{(n)}(r-1).
\end{equation}
Differentiation in $r$ maps $\delta^{(k)}$ to $\delta^{(k+1)}$, and $(r-1)\delta^{(k)}(r-1)=-k\delta^{(k-1)}(r-1)$. Thus, combining the coefficients $a_k$ into a vector, we have
\begin{equation}
\partial_r\rightarrow
\begin{pmatrix}
    0 & 0 & 0 & 0 & \ldots\\
    1 & 0 & 0 & 0 & \ldots\\
    0 & 1 & 0 & 0 & \ldots\\
    0 & 0 & 1 & 0 & \ldots\\
    \vdots & \vdots & \vdots & \vdots & \ddots\\
\end{pmatrix},\ 
r-1\rightarrow
\begin{pmatrix}
    0 & -1 & 0 & 0 & \ldots\\
    0 & 0 & -2 & 0 & \ldots\\
    0 & 0 & 0 & -3 & \ldots\\
    0 & 0 & 0 & 0 & \ldots\\
    \vdots & \vdots & \vdots & \vdots & \ddots\\
\end{pmatrix}.
\end{equation}
We then express the second order operator $P(\omega)^*$ acting on~\eqref{EqSumDelta} as an $N\times N$ matrix, $N=n+2$. Its nullspace consists of the sought-after dual states. As a consistency check, one can either verify the dual states by hand, or one can increase $N$ further and check that the subspace of the nullspace consisting of vectors with many trailing $0$'s is independent of $N$.

In Tables~\ref{D4m0DualStates}--\ref{D6m6DualStates}, we list a few examples of dual states supported at the dS horizon found in this manner. For brevity, we write
\begin{equation}
  \delta \equiv \delta(r-1),\quad
  \delta' \equiv \delta'(r-1),\quad\ldots
\end{equation}
The dual state $\Theta(1-r)$ for $l=0$, $\omega=0$ in $D=4$ and $D=6$ is not covered by the previous considerations, but can easily be verified by hand.

For $D=4$, we also list the corresponding mode solutions which can be found via expansion of~\eqref{EqQNMExp2} into spherical harmonics, or directly by solving~\eqref{EqQNMEq}. The relationship between mode solutions $u_j$ and $v_j$ is consistent with equation~\eqref{EqDualGuess2}, except that we normalized the dual states $v_j$ so that formulas~\eqref{EqDualcj} and \eqref{EqDualNorm} with $a_j=1$ compute the expansion coefficient of $u_j$ in the QNM expansion~\eqref{EqQNMExp1}. (The spherical harmonics $Y_{l m}$ are normalized so that $\iint_{\mathbb{S}^2} |Y_{l m}(\theta,\phi)|^2\,\sin\theta\,\dd\theta\,\dd\phi=1$.)

\begin{table}
    \centering
    \begin{tabular}{|c|c|c|c|}
        \hline
        \rowcolor[gray]{0.9}[\tabcolsep]
         \!$l$\! & \!$\omega$\! & \!mode sol.\ ($u_j$)\! & \!dual state ($v_j$)\! \\\hline
         \!0\! & \!0\! & \!$Y_{0 0}$\! & \!$Y_{0 0}\Theta(1-r)$\! \\\hline
         \!1\! & \!$-i$\! & \!$Y_{1 m}r$\! & \!$\tfrac13 Y_{1 m}\delta$\! \\\hline
         \!0\! & \!$-2 i$\! & \!$Y_{2 m}(r^2-3)$\! & \!$\tfrac{1}{6}Y_{0 m}(2\delta+\delta')$\! \\\hline
         \!2\! & \!$-2 i$\! & \!$Y_{2 m}r^2$\! & \!$\tfrac{1}{15}Y_{2 m}(\delta-\delta')$\! \\\hline
         \!0\! & \!$-3 i$\! & \!$Y_{0 0}$\! & \!$\tfrac13 Y_{0 0}(3\delta+3\delta'+\delta'')$\! \\\hline
         \!1\! & \!$-3 i$\! & \!$Y_{1 m}(r^3-5 r)$\! & \!$\tfrac{1}{30}Y_{1 m}(2\delta'+\delta'')$\! \\\hline
         \!3\! & \!$-3 i$\! & \!$Y_{3 m}r^3$\! & \!$\tfrac{1}{105}Y_{3 m}(-3\delta'+\delta'')$\! \\\hline
    \end{tabular}
    \caption{The first few normalized mode solutions ($u_j$) and dual states ($v_j$) for a massless scalar field in $D=4$.}
    \label{D4m0DualStates}
\end{table}

\begin{table}
    \centering
    \begin{tabular}{|c|c|c|c|}
        \hline
        \rowcolor[gray]{0.9}[\tabcolsep]
         \!$l$\! & \!$\omega$\! & \!\!mode sol.\ ($u_j$)\!\! & \!dual state ($v_j$)\! \\\hline
         \!0\! & \!$-i$\! & \!$Y_{0 0}$\! & \!$Y_{0 0}\delta$\! \\\hline
         \!0\! & \!$-2 i$\! & \!$Y_{0 0}$\! & \!$Y_{0 0}(\delta+\delta')$\! \\\hline
         \!1\! & \!$-2 i$\! & \!$Y_{1 m}r$\! & \!$-\tfrac13 Y_{1 m}\delta'$\! \\\hline
         \!0\! & \!$-3 i$\! & \!$Y_{0 0}(r^2+3)$\! & \!$\tfrac16 Y_{0 0}(2\delta+2\delta'+\delta'')$\! \\\hline
         \!1\! & \!$-3 i$\! & \!$Y_{1 m}r$\! & \!$-\tfrac13 Y_{1 m}(\delta'+\delta'')$\! \\\hline
         \!2\! & \!$-3 i$\! & \!$Y_{2 m}r^2$\! & \!$\tfrac{1}{15}Y_{2 m}(-\delta-\delta'+\delta'')$\! \\\hline
    \end{tabular}
    \caption{Normalized mode solutions and dual states for a scalar field with conformal mass $m=\sqrt{2}$ in $D=4$.}
    \label{D4m2DualStates}
\end{table}

\begin{table}
    \centering
    \begin{tabular}{|c|c|c|}
        \hline
        \rowcolor[gray]{0.9}[\tabcolsep]
         $l$ & $\omega$ & dual state \\\hline
         0 & 0 & $\Theta(1-r)$ \\\hline
         1 & $-i$ & $\delta$ \\\hline
         0 & $-2 i$ & $4\delta+\delta'$ \\\hline
         2 & $-2 i$ & $-\delta+\delta'$ \\\hline
         1 & $-3 i$ & $4\delta'+\delta''$ \\\hline
         3 & $-3 i$ & $-3\delta'+\delta''$ \\\hline
    \end{tabular}
    \caption{Dual states for a massless scalar field in $D=6$.}
    \label{D6m0DualStates}
\end{table}

\begin{table}
    \centering
    \begin{tabular}{|c|c|c|}
        \hline
        \rowcolor[gray]{0.9}[\tabcolsep]
        $l$ & $\omega$ & dual state \\\hline
         0 & $-2 i$ & $\delta+\delta'$ \\\hline
         0 & $-3 i$ & $3\delta+3\delta'+\delta''$ \\\hline
         1 & $-3 i$ & $\delta'+\delta''$ \\\hline
         0 & $-4 i$ & $12\delta+12\delta'+5\delta''+\delta'''$ \\\hline
         1 & $-4 i$ & $3\delta'+3\delta''+\delta'''$ \\\hline
         2 & $-4 i$ & $-3\delta-3\delta'+\delta'''$ \\\hline
    \end{tabular}
    \caption{Dual states for a scalar field with conformal mass $m=\sqrt{6}$ in $D=6$.}
    \label{D6m6DualStates}
\end{table}

%%%%%%%%%%%%%%%%%%%%%%%%%%%%%%%%%%%%%%%%%%%%%%%%%%
\subsection{CONNECTION WITH WAVE EVOLUTION}
\label{SWave}

Consider a QNM $\omega_j$ so that the associated dual resonant state $v_j$ is supported on the dS horizon $r=L$. When the forcing $f$ or the initial data $(\Phi_0,\Phi_1)$ vanish near $r=L$, then the corresponding mode solution $u_j$ never contributes to the late-time asymptotics of the field $\Phi$. This is the setup of the wave evolution in \cite[\S IV.D]{BradyChambersLaarakeersPoissonSdSFalloff}.

This is further connected to the sharp Huygens principle \cite{WuenschHuygens,YagdjianHuygens,YagdjianHuygensDirac}: in the cases where \emph{all} dual resonant states are supported on the horizon, $\Phi$ in~\eqref{EqQNMExp1} decays faster than any exponential in $t_*$, (and in fact can be shown to vanish inside the horizon for late times). The converse is true as well: if the sharp Huygens principle holds, then all dual resonant states are supported on $r=L$. The validity of the sharp Huygens principle for even spacetime dimensions $D\geq 4$ and for the conformal mass $m=\frac12\sqrt{D(D-2)}$ follows directly from the fact that by~\eqref{EqKG},
\begin{equation}
\label{EqMinkEq}
  \tau^{-1-D/2}L^2(\Box-m^2)\Phi = (-\pa_\tau^2+\Delta_X)(\tau^{1-D/2}\Phi) = 0
\end{equation}
is the wave equation on $D$-dimensional Minkowski space, which satisfies the sharp Huygens principle.

\begin{figure}
    \centering
    \includegraphics[width=0.45\textwidth]{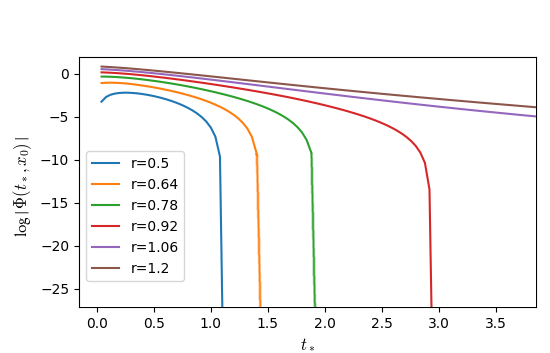}
    \caption{Logarithm of the value of solutions of the conformal wave equation in $D=4$ at a point $x_0$ with $|x_0|<L=1$. The initial data are $(0,(r^2-|\mathbf{x}|^2)\Theta(r^2-|\mathbf{x}|^2))$ for various values of $r$.}
    \label{logplot}
\end{figure}

Figure~\ref{logplot}, which concerns the conformally coupled scalar field on $D=4$ dS, demonstrates this phenomenon. (The solutions shown in Figure~\ref{logplot} were obtained by solving the initial value problem for $\tau^{1-D/2}\Phi$ on $D=4$ Minkowski space---see~\eqref{EqMinkEq}---using the well-known explicit solution formula, with numerical evaluation of the relevant integral over the base of the past light cone.) For initial data which are supported in $r<L=1$, the conformally coupled scalar field on $D=4$ dS decays to $0$ superexponentially fast in $r<1$ as $t_*\to\infty$, and in fact vanishes for sufficiently late times. By contrast, when the initial data are nonzero near $r=1$, the rate of decay of the amplitude of the field is exponential, and indeed $\sim e^{-t_*}$; this is consistent with the fact that the dominant QNM is $\omega=-i$, see Table~\ref{D4m2DualStates}. Therefore, Figure~\ref{logplot} demonstrates that the exceptional QNMs---which in this case constitute the \emph{full} set of QNMs---that were discarded in \cite{BradyChambersLaarakeersPoissonSdSFalloff} \emph{must be kept} and do contribute to the late time asymptotics when the initial data of the field are nonzero near the dS horizon. In particular, if one allows for such general initial data, a qualitative change in the late time asymptotics when the scalar field mass approaches special values (such as the conformal mass) observed in \cite[\S IV.D]{BradyChambersLaarakeersPoissonSdSFalloff} does \emph{not} take place.

\begin{figure}
  \centering
  \includegraphics[width=0.42\textwidth]{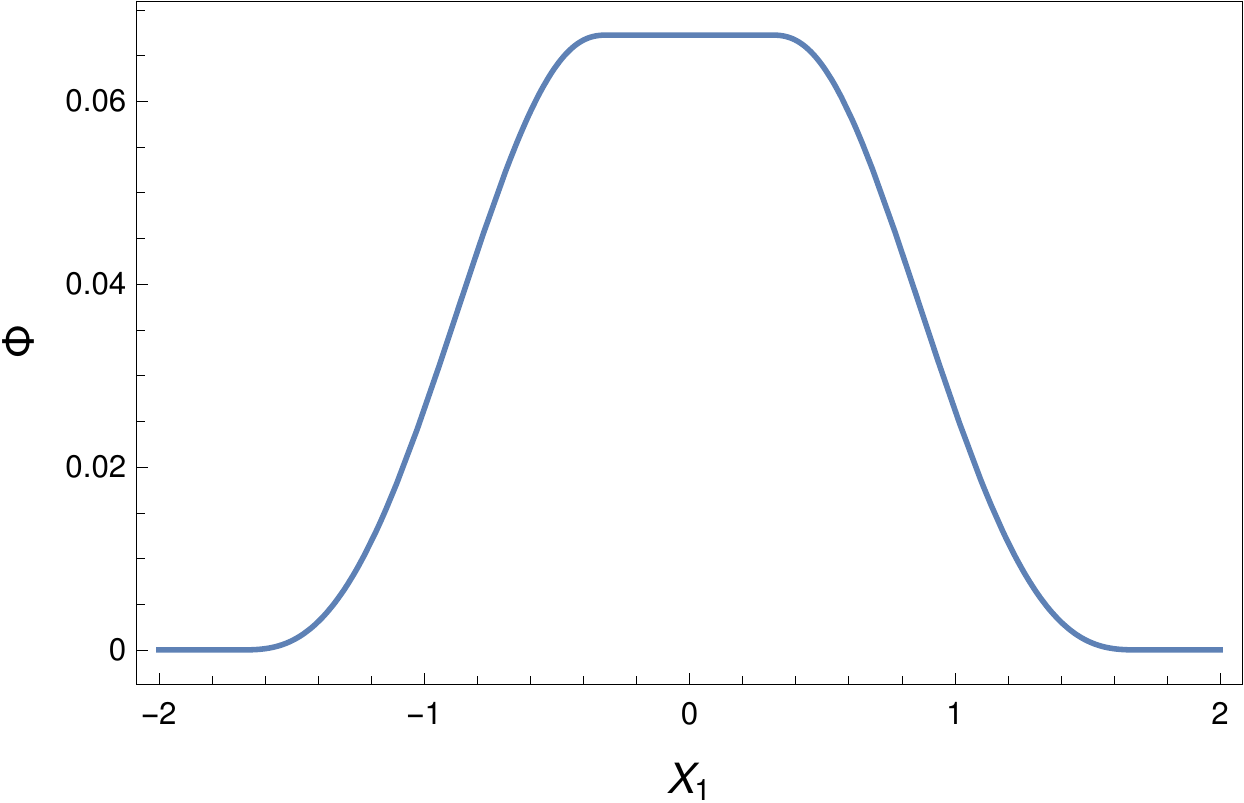}
  \caption{Amplitude of a real massless scalar field $\Phi(t_*,X)$ at points $X=(X_1,0,0)$, with initial data $(0,-(0.49-|\mathbf{x}|^2)\Theta(0.49-|\mathbf{x}|^2))$, after evolving for a long time in the coordinates $(t_*,X)$ from~\eqref{EqtauX}--\eqref{EqConf}.}
  \label{WithinBound}
\end{figure}

In some cases such as $D=2,m=0$ or $D=4,m=0$, only a single dual resonant state (at frequency $\omega=0$) is nonzero in $r<L$, thus $\Phi|_{r<L}$ is equal to a constant\footnote{Constant functions are the mode solutions at frequency $\omega=0$.} (and the energy density is zero) for late times when the initial data vanish near the horizon. This is closely related to the \emph{incomplete Huygens principle} \cite{YagdjianHuygens}. Figure~\ref{WithinBound} provides a perspective on this phenomenon from the perspective of the global dS spacetime~\eqref{EqConf} for $D=4,m=0$ and $L=1$: the scalar field $\Phi$, given by~\eqref{EqPhi}, \eqref{EqAnsatz}, \eqref{EqAnsatzSol} with $\lambda_-(0)=0$ and $\lambda_+(0)=3$, is plotted at a time $t_*\gg 1$ (i.e.\ $0<\tau\ll 1$); thus, the plot shows a very precise approximation of the asymptotic datum $u_-^{(0)}$. The flat middle segment is independent of $t_*$ in this regime. Hence, in coordinates $(t_*,x)$ adapted to the static patch as in~\eqref{EqtauX} (which zoom in exponentially fast in time around $X=0$), the scalar field is constant for large $t_*$.

\begin{figure}
  \centering
  \includegraphics[width=0.42\textwidth]{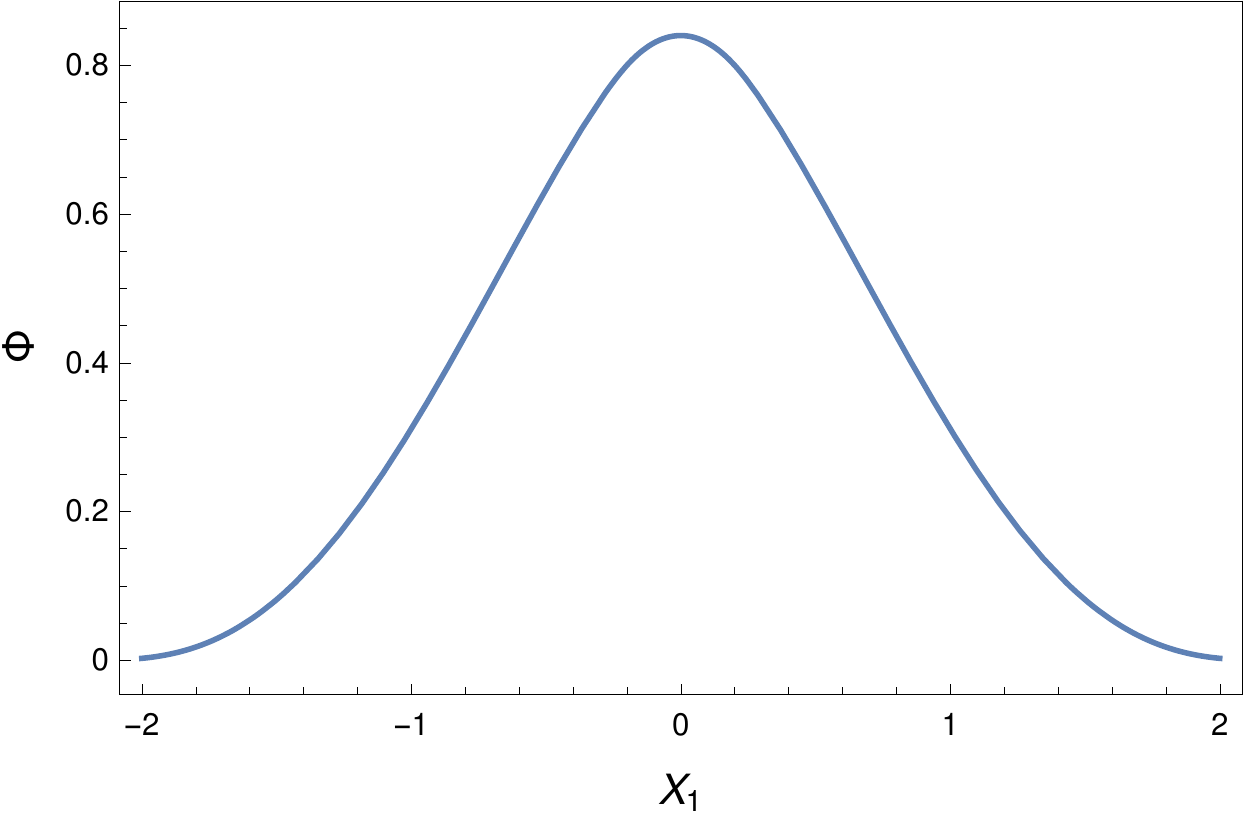}
  \caption{Amplitude of a real massless scalar field $\Phi(t_*,X)$ with initial data $(0,-(1.44-|\mathbf{x}|^2)\Theta(1.44-|\mathbf{x}|^2))$, after evolving for a long time in the coordinates $(t_*,X)$.}
  \label{OnBound}
\end{figure}

By contrast, Figure~\ref{OnBound} shows $\Phi(t_*,X)$ for $t_*\gg 1$ for initial data which are nontrivial near the dS horizon. Again $\Phi(t_*,X)$ is independent of $t_*$ up to error terms of size $e^{-2 t_*}$, hence the Figure approximately shows $u_-^{(0)}$. Since therefore $u_-^{(0)}$ is not constant near $X=0$, switching back to $(t_*,x)$ coordinates reveals that $\Phi(t_*,x)$ approaches a constant value exponentially fast in time, with the exponential rate of convergence an integer multiple of $\kappa$.

Figures~\ref{WithinBound}--\ref{OnBound} were obtained via numerical evaluation of an analytic representation formula: by finite speed of propagation, one may replace the spatial manifold $\R^D$ by a large torus. Upon expanding the field $\Phi(t_*,X)$ into Fourier modes $e^{i X\cdot k}$ in $X$ (where $k$ lies on an appropriate $D$-dimensional lattice), one obtains from~\eqref{EqKG} (with $D=4$ and $m=0$) an infinite family of ODEs $(-(\tau\pa_\tau)^2+3\tau\pa_\tau-\tau^2|k|^2)\Phi_k(\tau)=0$. Each ODE has an explicit solution which is a linear combination of $\cos(|k|\tau)+|k|\tau\sin(|k|\tau)$ and $-|k|\tau\cos(|k|\tau)+\sin(|k|\tau)$; the coefficients are determined from the initial conditions. Fourier inversion and evaluation at $\tau\ll 1$ produces the figures.

For other values of the scalar field mass or spacetime dimension, and especially if one is interested in the precise behavior of the field in or near the static patch as $t_*\to\infty$, a full numerical evolution scheme (e.g.\ as in \cite{BradyChambersLaarakeersPoissonSdSFalloff}) may be needed. The above special choices of $D$ and $m$ allow us to avoid this; and in any case they demonstrate most clearly our main arguments.

%%%%%%%%%%%%%%%%%%%%%%%%%%%%%%%%%%%%%%%%%%%%%%%%%%
\subsection{CONNECTION WITH THE GREEN'S FUNCTION IN STATIC COORDINATES}
\label{SGreen}
%%%%%%%%%%%%%%%%%%%%%%%%%%%%%%%%%%%%%%%%%%%%%%%%%%

The Green's function on dS in frequency space was explicitly computed in static coordinates~\eqref{EqdSStatic} in \cite{BradyChambersLaarakeersPoissonSdSFalloff,ChoudhuryPadmanabhanQNMforSdS,NatarioSchiappaQNM}. This amounts to constructing an inverse $G^s(\omega;x,x')$ of the spectral family in static coordinates, $P^s_m(\omega) = e^{i\omega t}(\Box-m^2)e^{-i\omega t}$ (see~\eqref{EqSpectralFy} for the explicit expression), with the requirement that $G^s(\omega;x,x')$ (defined for $|x|,|x'|<L$) be outgoing as $|x|\to L$ and regular at $x=0$. QNMs were then defined as the poles $\omega$ of $G^s(\omega;x,x')$. Since
\begin{equation}
  G^s(\omega) = e^{-i\omega(t_*-t)}G(\omega)e^{i\omega(t_*-t)} \sim e^{i\omega r_*}G(\omega)e^{-i\omega r_*},
\end{equation}
the pole of $G$ in~\eqref{EqGreensPole} does not survive upon restricting $x,x'$ to the static patch $|x|,|x'|<L$ precisely when the dual resonant state $v_j(x')$ vanishes in $|x'|<L$, i.e.\ when it is supported on the dS horizon.

Thus, the set of poles of $G^s$ can be strictly smaller than that of $G$, with explicit instances where this happens given in \S\ref{SExpl}. This is the reason for the contradictory statements regarding the existence of QNMs in the literature when they are defined in terms of the poles of the meromorphically continued Green's function in static coordinates. It would be interesting to derive an explicit formula for the Green's function $G(\omega;x,x')$ in coordinates which are regular across the dS horizon, though such a construction is necessarily quite subtle since $G(\omega;x,x')$ is a very singular distribution at $|x'|=L$.

%%%%%%%%%%%%%%%%%%%%%%%%%%%%%%%%%%%%%%%%%%%%%%%%%%%%%%%%%%%%%%%%%%%%%%
\section{CONCLUSIONS AND OUTLOOK}

We have conclusively shown the existence of QNMs for massive scalar fields on dS, and produced an explicit formula for the amplitude with which any QNM and mode solution appears in the QNM expansion of the field.

We have moreover produced explicit formulas which demonstrate that the QNMs as well as the mode solutions depend continuously on the mass $m$ of the scalar field. One can moreover show that also the dual resonant states depend continuously (as distributions, i.e.\ when paired against any fixed test function) on $m$. In particular, the qualitative changes suggested in \cite{BradyChambersLaarakeersPoissonSdSFalloff} when $m$ tends to special values (such as $0$ or the conformal mass) do, in fact, \emph{not} take place for general initial data which are allowed to be nonzero near the dS horizon. Our analysis gives the correct late time behavior for \emph{all} initial conditions of the scalar field: while prior results, such as those in \cite{BradyChambersLaarakeersPoissonSdSFalloff}, suggested a very rapidly decaying late time tail of the field for special values of $m$, we have demonstrated that the decay rate is, in fact, given in terms of an explicit QNM, though the amplitude of the corresponding mode solution in the QNM expansion of the field vanishes for initial data which vanish near the dS horizon. Note however that there is no a priori reason why the scalar field (or other fields of interest, see below) should initially vanish near the horizon. In other words, we are able to explain the correct (numerically observed) late time tails, and can moreover do so entirely within the framework of QNM expansions.

As another setting in which the QNMs of dS play an important role, we mention the QNM spectrum of SdS black holes as the black hole mass $M_\bullet$ tends to $0$. This is the subject of \cite{HintzXieSdS}, where it is shown that the QNM spectrum tends to that of dS in any bounded subset of the complex plane. Again, it is crucial to keep also those dS QNMs whose dual resonant states are supported on the dS horizon, as otherwise some SdS QNMs would disappear in the limit $M_\bullet\to 0$.

On pure dS, our method for calculating QNMs and mode solutions can be generalized to many other equations, such as the Maxwell and linearized Einstein equations; related results appear in \cite[\S4.1]{HintzVasyKdsFormResonances}, \cite[Appendix~C]{HintzVasyKdSStability}. On the other hand, we do not have an equally efficient method for the calculation of dual resonant states in such general settings at this time; we only obtained explicit formulas~\eqref{EqDualGuess}--\eqref{EqDualGuess2} in the scalar case.

We end by suggesting an intriguing potential application of dual resonant states on black hole spacetimes, namely that via their connection to the coefficients in QNM expansions---which can be experimentally measured \cite{IsiGieslerFarrScheelTeukolskyNoHair}---they may give useful information about the conditions close to the black hole from far field gravitational wave measurements. The key observation here is that, just like the mode solutions themselves, the dual states corresponding to QNMs with large real part (and small imaginary part) are localized near the photon sphere, hence the coefficients of the QNM expansion give averaged information on initial conditions there.

\bigskip

%%%%%%%%%%%%%%%%%%%%%%%%%%%%%%%%%%%
\noindent{\bf{\em Acknowledgments.}} %
%%%%%%%%%%%%%%%%%%%%%%%%%%%%%%%%%%%
%
The authors are very grateful to Vitor Cardoso and an anonymous referee for helpful comments and suggestions. PH gratefully acknowledges support from the NSF under Grant No.\ DMS-1955614 and from a Sloan Research Fellowship. Part of this research was conducted during the time PH served as a Clay Research Fellow. YX would like to thank the MIT UROP office for this opportunity and gratefully acknowledges support from the Reed Fund, Baruch Fund, and Anonymous Fund.

\bigskip
\appendix
%%%%%%%%%%%%%%%%%%%%%%%%%%%%%%%%%%%
\section{CONVERGENCE OF~\eqref{EqAnsatz} FOR ANALYTIC DATA}
\label{AppConv}
%%%%%%%%%%%%%%%%%%%%%%%%%%%%%%%%%%%

We shall need a general estimate, adapted from \cite[\S2]{SjostrandMicrolocal}, for bounding powers of the Laplacian applied to real-analytic functions. For $R_0>0$, denote $B(R_0)=\{X\in\R^D\colon|X|<R_0\}$ and $\tilde B(R_0)=\{\lambda X\colon X\in B(R_0),\lambda\in\C,|\lambda|=1\}$. If $u=u(X)$ is real-analytic in $X\in\R^D$, and extends to a holomorphic function of $X\in\C^D$ in a neighborhood of $\tilde B(R_0)$ for some $R_0>0$,\footnote{This is always true for some sufficiently small $R_0>0$ by definition of real-analyticity.} then for all $j=0,1,2,\ldots$, we have
\begin{equation}
\label{EqAppEst}
  |\Delta_X^j u(0)| \leq C(D) R_0^{-2 j} 2^{2 j} j!^2 (j+1)^{\frac{D}{2}-1} \sup_{\tilde B(R_0)} |u|.
\end{equation}
The proof will be given below. In the context of~\eqref{EqAnsatz}--\eqref{EqAnsatzSol}, we apply~\eqref{EqAppEst} to $u=u_\pm^{(0)}$. Further note that since $p(\lambda)\sim\lambda^2$ for large $\lambda$, there exists, for any $\theta\in(0,1)$ (and with the scalar field mass $m$ fixed), an integer $k_0\geq 1$ so that for all $k\geq k_0$ one has $|p(\lambda_\pm(m)+2 k)|\geq\theta(2 k)^2$, and therefore
\begin{equation}
  |q_\pm^{(j)}| \leq C(\theta) \theta^{-j} (2^j j!)^{-2}.
\end{equation}
This implies
\begin{equation}
  |u_\pm^{(j)}(0)| \leq C(D,\theta)(R_0\theta^{1/2})^{-2 j} (j+1)^{\frac{D}{2}-1} \sup_{\tilde B(R_0)}|u|,
\end{equation}
and therefore the series
\begin{equation}
  u_\pm(\tau,0) = \sum_{j=0}^\infty \tau^{2 j}u_\pm^{(j)}(0)
\end{equation}
converges absolutely for $|\tau|<R_0$ since $\theta<1$ was arbitrary. Working at general points $X\in\R^D$, we have shown that the series~\eqref{EqAnsatz} converges absolutely and defines a real-analytic function of $(\tau,X)$ in a neighborhood of $\tau=0$, as claimed.

We now turn to the proof of~\eqref{EqAppEst}. By passing to $\tilde X=X/R_0$ and thus $\Delta_X=R_0^{-2}\Delta_{\tilde X}$, we may assume that $R_0=1$. Moreover, dividing $u$ by $\sup_{\tilde B(1)}|u|$ (unless $u$ is identically $0$, in which case~\eqref{EqAppEst} is trivial), we may assume $\sup_{\tilde B(1)}|u|=1$. Note moreover that~\eqref{EqAppEst} is trivial for $j=0$, hence we only consider $j\geq 1$.

To proceed, we note that when $p=p(X)$ is a homogeneous polynomial of degree $2 j$, then $X\cdot\pa_X p=2 j p$ and thus
\begin{equation}
\begin{split}
  &\int_{\R^D} e^{-X^2/2}p(X)\,\dd X \\
  &\quad = \frac{1}{2 j}\int_{\R^D} -\pa_X\cdot(X e^{-X^2/2})p(X)\,\dd X \\
  &\quad = \frac{1}{2 j} \int_{\R^D} \Delta_X(e^{-X^2/2}) p(X)\,\dd X \\
  &\quad = \frac{1}{2 j} \int_{\R^D} e^{-X^2/2}\Delta_X p(X)\,\dd X.
\end{split}
\end{equation}
Since $\Delta_X p(X)$ is homogeneous of degree $2(j-1)$, we can proceed inductively and obtain
\begin{equation}
\label{EqAppIdentity}
  \int_{\R^D} e^{-X^2/2}p(X)\,\dd X = (2\pi)^{D/2}\frac{\Delta_X^j p}{j! 2^j},
\end{equation}
with $\Delta_X^j p$ a constant. Applying this to the polynomial $p(X)=\sum_{|\alpha|=2 j}\frac{\pa_X^\alpha u(0)}{\alpha!}X^\alpha$ (with $\Delta_X^j p=\Delta_X^j u(0)$) gives
\begin{equation}
\label{EqAppIM}
  |\Delta_X^j u(0)| \leq (2\pi)^{-D/2}j! 2^j \int_{\R^D} e^{-X^2/2}|p(X)|\,\dd X.
\end{equation}
Fixing any unit vector $\hat X\in\R^D$, consider $u_{\hat X}(z)=u(z\hat X)$ as a function of the single complex number $z\in\C$, defined in a neighborhood of $|z|\leq 1$. Then the Cauchy integral formula gives $|\pa_z^{2 j}u_{\hat X}(0)| \leq (2 j)! \sup_{|z|\leq 1} |u_{\hat X}(z)|$ and therefore
\begin{equation}
  |p(\hat X)| = \frac{\pa_z^{2 j}u_{\hat X}(0)}{(2 j)!} \leq \sup_{\tilde B(1)} |u| = 1.
\end{equation}
Since $p$ is homogeneous of degree $2 j$, we conclude from this and~\eqref{EqAppIM}, and using the identity~\eqref{EqAppIdentity} to the polynomial $|X|^{2 j}$, that
\begin{equation}
\label{EqAppEst2}
\begin{split}
  |\Delta_X^j u(0)| &\leq (2\pi)^{-D/2}j! 2^j \int e^{-X^2/2}|X|^{2 j}\,\dd X \\
    &= \Delta_X^j|X|^{2 j}.
\end{split}
\end{equation}
This can be computed in the radial coordinate $R=|X|$ and equals $(R^{-D+1}\pa_R R^{D-1}\pa_R)^j R^{2 j}$, which evaluates to
\begin{equation}
\begin{split}
  &(2 j)(2 j+D-2)\cdot(2 j-2)(2 j+D-4)\cdots 2\cdot D \\
  &\ = 2^{2 j} j!(j-1)! \frac{(j-1+\frac{D}{2})\cdots(1+\frac{D}{2})}{(j-1)\cdots 1}\cdot\frac{D}{2} \\
  &\ \leq C(D)2^{2 j}j!(j-1)! j^{D/2}\frac{D}{2}.
\end{split}
\end{equation}
(The final inequality follows from Stirling's formula, since the fraction is equal to $\frac{\Gamma(j+D/2)}{\Gamma(j)\Gamma(D/2)}$.) Plugging this into~\eqref{EqAppEst2} and simplifying finishes the proof of~\eqref{EqAppEst}.

%\bibliography{/home/peter/research/bib/math.bib,/home/peter/research/bib/mathcheck.bib,/home/peter/research/bib/phys.bib}
\input{dsqnm.bbl}

\end{document}